\documentclass[a4paper,11pt]{article}
\usepackage{amsmath,amssymb,mathtools}
\usepackage{color}
\usepackage{graphicx}
\usepackage{subfigure}
\usepackage{hyperref}
\usepackage{mathrsfs}
\usepackage{amsthm}
\usepackage{float} 
\usepackage{ulem}
\usepackage[text={17cm,24.5cm},centering]{geometry}
\def \be {\begin{equation}}
\def \ee {\end{equation}}
\def \ba {\begin{array}}
\def \ea {\end{array}}
\def \bea{\begin{eqnarray}}
\def \eea{\end{eqnarray}}

\def \mA {\mathcal A}
\def \mB {\mathcal B}

\def \mD {\mathcal D}
\def \mO {\mathcal O}
\def \mPO{\mathcal{P}_{\mathcal{O}}}
\def \mX {\mathcal{X}}
\def \mY {\mathcal{Y}}

\begin{document}

\title{Entanglement spectrum of geometric states}
\author{Wu-zhong Guo\footnote{wuzhong@hust.edu.cn}~}

\date{}
\maketitle

\vspace{-10mm}
\begin{center}
{\it School of Physics, Huazhong University of Science and Technology,\\
 Wuhan, Hubei
430074, China
\vspace{1mm}
}
\vspace{10mm}
\end{center}

\begin{abstract}
 The reduced density matrix of a given subsystem, denoted by $\rho_A$, contains the information on subregion duality in a holographic theory.  We may extract the information by using the spectrum (eigenvalue) of the matrix, called entanglement spectrum in this paper.  We evaluate  the density of eigenstates, one-point and two-point correlation functions  in the microcanonical ensemble state $\rho_{A,m}$ associated with an eigenvalue $\lambda$ for some examples, including a single interval and two intervals in vacuum state of 2D CFTs.  We find there exists a microcanonical ensemble state with $\lambda_0$  which can be seen as an approximate state of $\rho_A$. The parameter $\lambda_0$ is obtained in the two examples.  For a general geometric state, the  approximate microcanonical ensemble state also exists. The parameter $\lambda_0$ is associated with the entanglement entropy of $A$ and R\'enyi entropy in the limit $n\to \infty$.   As an application of the above conclusion we reform the equality case of the Araki-Lieb inequality of the entanglement entropies of  two intervals in vacuum state of 2D CFTs  as conditions of Holevo information. We show the constraints on the eigenstates.  Finally, we point out some unsolved problems and their significance on understanding the geometric states.
\end{abstract}

\newpage

\section{Introduction}
For a quantum field theory with a gravity dual, there exists some states that can be effectivelly described by a classical geometry in the semi-classical limit $G\to 0$.  We call these states geometric states.  These states should show special properties, for example the probes, such as correlation functions, in geometric states should be well-defined in the limit $G\to 0$ .  Among all the probes there exists some special ones that can be associated with geometric objects in the bulk.  We may call them  geometric probes.\\
At present we still have no methods to judge whether a state is geometric or not. But we can use the geometric probes to detect the properties of geometric states. Quantum entanglement is a useful concept to understand the geometry.  Among various measures of entanglement, entanglement entropy (EE) of a subsystem $A$ is the most useful one since the discovery  of the Ryu-Takayanagi (RT) formula \cite{Ryu:2006bv}
\bea\label{RT}
S_A=\frac{\text{min}(\text{Area}\{\gamma_A\})}{4G}, 
\eea 
where $S_A$ is the EE of the subsystem $A$, $\gamma_A$  is the bulk surface that is homologous to $A$.  By using this simple relation one may obtain fruitful results which help us to catch the properties of the geometric states. Motivated by the holographic EE, in  \cite{VanRaamsdonk:2009ar}\cite{VanRaamsdonk:2010pw}the author finds the secret relation between quantum entanglement and connectivity  of spacetime.   Another way to understand geometric states is using  the analogy between tensor networks and the geometry of spatial slice of AdS/CFT\cite{Swingle:2009bg}.  The geometric states can also be constrained by using the inequality of entropy. In \cite{Hayden:2011ag} the authors show the so-called tripartite information (a linear combinations of EEs of arbitrary three regions) in the  geometric states should be non-negative, thus give strong constraints on the geometric states. Direct construction of geometric states as coherent state are studied in \cite{Botta-Cantcheff:2015sav}-\cite{Marolf:2017kvq}, see also \cite{Belin:2018fxe}-\cite{Arias:2020qpg}. On the field theory side we find a series of necessary conditions on expectation values of quasi-primary operators in AdS$_3$/CFT$_2$ by using the scaling behavior of large $c$ limit\cite{Guo:2018fnv}.  \\
Here we only list some attempts to characterize various  aspects of geometric states.  Many of them are associated with EE. 
But EE as a functional of the reduced density matrix $\rho_A$ only contains limited information of $\rho_A$.  In this paper one of our motivation is to study the spectrum of $\rho_A$ for the geometric states.  We will shortly call them entanglement spectrum following \cite{Li2008}, in which it is  used to identify topological order.  In this paper we will also use the term ``eigenvalue'' instead of spectrum, though in general they may be different. \\
Generally, the density matrix $\rho_A$ has the following spectrum decomposition
\bea\label{spectrumdecompositionintroduce}
\rho_{A}=\sum_i \lambda_i |\lambda_i\rangle_A ~_A\langle \lambda_i|,
\eea
where $|\lambda_i\rangle_A$  are generally degenerate. The maximal eigenvlue of $\rho_A$ is $\lambda_m$. We can parameterize the eigenvalue $\lambda$ as $\lambda=\lambda_m e^{-t}$ with $t\in [0,+\infty)$. The distributions of entanglement spectrum can be obtained if one knows the R\'enyi entropy $S^{(n)}_A$ of all the index $n$. This is done in \cite{Calabrese2008} for a single interval in the vacuum state of 2D CFTs.  The results can be easily generalized to the cases that are studied in \cite{Cardy:2016fqc}.  In this paper we will discuss more non-trivial examples.  Further, by using the method of inverse Laplace transformation similar with \cite{Kraus:2016nwo},\cite{Romero-Bermudez:2018dim}\cite{Brehm:2018ipf}, we also study the correlation functions in the so-called microcanonical ensemble states $\rho_{A,m}$, 
\bea
\rho_{A,m}:= \frac{1}{\mathcal{P}(t)}\sum_i|\lambda_i\rangle_A~_A\langle \lambda_i| \delta(t_i-t),
\eea
where $\mathcal{P}(t)$ is the density of eigenstate with respect to $t$.
Our motivation is to explore the behavior of eigenvalues and eigenstates of $\rho_A$ in the large $c$  limit in 2D CFTs.  Our examples include a single interval and  two intervals in vacuum state. The examples show  one could find a microcanonical ensemble state to be an approximate state of $\rho_A$ in the large $c$ limit, if the probes are located in a small region of $A$ and far away from the boundary of $A$.  We make the above conclusions by directly comparing the one-point and two-point functions in $\rho_{A,m}$  with the ones in $\rho_A$. \\
The results can be generalized to arbitrary geometric states by using the fact that $S^{(n)}_A\sim O(c)$ or $O(1/G)$, which follows the holographic R\'enyi entropy proposal in \cite{Dong:2016fnf}.  The microcanonical ensemble state are associated with the EE $S(\rho_A)$ and $S^{\infty}=\lim_{n\to \infty}S^{(n)}_A$. More preciesely,  the parameter $t_0$ of the special microcanonical ensemble state is given by
\bea
t_0 =S(\rho_A)-S^{\infty}.
\eea 
Our results actually give a general feature of the entanglement spectrum of geometric states. \\
One of the interesting phenomenon that one  can obtain  from (\ref{RT}) is the phase transition of the EE of two disconnected subsystems. Without loss of generality we will  focus on two intervals in the vacuum state of 2D CFTs. In this paper we will choose $A_1=[-R,-T]$ and $A_2=[T,R]$ with $R>T>0$ and denote $A_3=[-T,T]$.  By using the RT formula (\ref{RT}) one can find the critical point at $T_0=(3-2\sqrt{2})R$. For $T>T_0$ we have $S_{A_1A_2}=S_{A_1}+S_{A_2}$ which means $A_1$ and $A_2$ lose correlations. For the case $T<T_0$  we have 
\bea\label{ArakiLiebequality}
S_{A_1A_2}=S_{A_3}+S_{A_1A_2A_3}.
\eea
The equality is satisfied for the  Araki-Lieb inequality $S_{AB}\ge S_{A}-S_B$  with $A=A_1A_2$ and $B=A_3$. \\
In this paper we also study the implication of the equality condition of Araki-Lieb inequality on the spectrum of the reduced density matrices $\rho_{A_1A_2A_3} $ and $\rho_{A_3}$.  This is based on reforming the above condition as Holevo information, that is
\bea
\chi(\rho_{A_1A_2})=H(\lambda_i)+O(c^0) \quad \text{and}\quad \chi(\rho_{A_3})=0+O(c^0),
\eea
where $H(\lambda_i)$ is the EE of $A$, $\rho_{A_1A_2}=\sum_i \lambda_i \rho_{i,A_1A_2}$ and $\rho_{A_3}=\sum_i \lambda_i \rho_{i,A_3}$ with 
 $\rho_{i,A_1A_2}:=tr_{A_3} |\lambda_i\rangle_A~_A\langle \lambda_i|$ and $\rho_{i,A_3}:= tr_{A_1A_2} |\lambda_i\rangle_A~_A\langle \lambda_i|$.
This condition gives the constraints on measurement or quasi-primary operators of the vacuum conformal family in the single eigenstate. \\
Finally, we discuss the possible extensions based on present paper. One interesing question is how to find the critical point of the distinguishability and indistinguishability of $\rho_{A,m}$ with $t=t_0$ from $\rho_A$. We also discuss the possibility of  the Holevo information as  a geometric probe.  At last we compare the microcanonical ensemble state in this paper with the approximate state constructed by tensor networks and the fixed-area state in the quantum error-correcting code of AdS.
\section{Entanglement spectrum and microcanonical ensemble state}
\subsection{Density of eigenstates }\label{densitysingle}
For a pure state $|\psi\rangle$ a Schmidt decomposition of the subsystem $A$ and its complement $\bar A$ is given by 
\bea
|\psi\rangle=\sum_i \sqrt{\lambda_i} |\lambda_i\rangle_A \otimes |\bar{\lambda}_i\rangle_{\bar A},
\eea
where $|\lambda_i\rangle_{A(\bar A)}\in \mathcal{H}_{A(\bar A)}$ satisfy $~_{A(\bar A)}\langle \lambda_i| \lambda_j\rangle_{A(\bar A)}=\delta_{ij}$. It is obvious that $|\lambda_i\rangle_{A(\bar A)}$ are the eigenstates of $\rho_{A(\bar A)}$.  For QFTs the spectrum of $\rho_A$ is continous, thus the sum of the decomposition should be replaced by integration.  If $|\psi\rangle=|0\rangle$ and $A$ is an interval,  the eigenvalues and the corresponding eigenstates should only depend on  the length of the interval by translation invariance. The reduced density matrix of $A$ is $\rho_{A}=\sum_i \lambda_i |\lambda_i\rangle_A ~_A\langle \lambda_i|$, where $\lambda_i$ is the eigenvalue of $\rho_A$.  The EE of $A$ is given by the Shannon entropy $H(\lambda_i):= -\sum_i \lambda_i \log \lambda_i$.\\
With the R\'enyi entropy  $S^{(n)}_A:= \frac{\log tr \rho^n_A}{1-n}$
of all the index $n$, one could construct density of eigenstates of the reduced density matrix $\rho_A$. The density of the states at the eigenvalues $\lambda$ is defined as 
\bea
P(\lambda):= \sum_k \delta(\lambda_k -\lambda),
\eea
which satisfies the normalization condition  $\int_{0}^{\lambda_m} d\lambda \lambda P(\lambda)=1$
with $\lambda_m$ is the maximal eigenvalue of $\rho_A$.  One may calculate $P(\lambda)$ of one interval in vacuum state of 2D CFTs by using the method in \cite{Calabrese2008}. For our motivation we will use the inverse Laplace transformation method that is mentioned in the same reference. The Laplace transformation method is used in \cite{Hung:2011nu} to study the entanglement sepectrum of a sphere in the vacuum state.  By the definition of the R\'enyi entropy we can find the maximal eigenvalue $\lambda_m$ of $\rho_A$ by
\bea
b:=-\log \lambda_m=\lim_{n\to \infty}S^{(n)}_A.
\eea
We also have
\bea
tr \rho_A^n =\int_{0}^{\lambda_m} d\lambda \lambda^n P(\lambda)=e^{(1-n)S^{(n)}_A}.
\eea
Let's parameterize $\lambda$  as $\lambda=\lambda_m e^{-t}$, the integral becomes the form of Laplace transformation, 
\bea
\int_0^{+\infty}P(\lambda_m e^{-t}) e^{-(n+1)t}dt =\lambda_m^{-(n+1)}e^{(1-n)S^{(n)}_A}.
\eea
By inverse Laplace transform we have the density of eigenstates with respect to $t$
\bea\label{distributiont}
\mathcal{P}(t):=P(\lambda_m e^{-t}) \lambda_m e^{-t}=\mathcal{L}^{-1}\left[ \lambda_m^{-n}e^{(1-n)S^{(n)}_A}\right](t):=\frac{1}{2\pi i}\int_{\gamma-i\infty}^{\gamma+i\infty}dn e^{n t}e^{nb+(1-n)S^{(n)}_A} .
\eea
Using this one can get the density of eigenstates $P(\lambda)$
\bea
P(\lambda) =\frac{\lambda_m}{\lambda}\mathcal{L}^{-1}\big[ \lambda_m^{-(n+1)}e^{(1-n)S^{(n)}_A}\big](t)|_{t\to\log \frac{\lambda_m}{\lambda}}.
\eea
\subsubsection{One-point functions}
The one-point correlation functions are useful probes to study the properties of the eigenstates of $\rho_A$. For an operator $\mathcal{O}$ we define the one-point functions 
\bea
P_{\mathcal{O}}(\lambda):= \sum_i ~_A\langle \lambda_i | \mathcal{O}|\lambda_i \rangle_A \delta(\lambda_i-\lambda).
\eea
The density of eigenstate $P(\lambda)$ is a special case with $\mathcal{O}=I$.  We assume the operator $\mathcal{O}$ is located in the region $A$. So we have 
\bea 
tr(\rho_A \mathcal{O}) =\sum_j ~_A\langle \lambda_j| \mathcal{O}|\lambda_j\rangle_A=\int_0^{\lambda_m}d\lambda \lambda P_{\mathcal{O}}(\lambda),
\eea
and
\bea
tr (\rho_A^n \mathcal{O}):= \int_0^{\lambda_m}d\lambda \lambda^n P_{\mathcal{O}}(\lambda).
\eea
Taking $\lambda=\lambda_m e^{-t}$, the above integral becomes 
\bea
tr (\rho_A^n \mathcal{O})= \lambda_m^{n+1}\int_0^{+\infty}dt e^{-(n+1)t} P_{\mathcal{O}}(\lambda_m e^{-t}).
\eea
By an inverse Laplace transformation we get
\bea\label{key0}
\mathcal{P}_{\mathcal{O}}(t):=  P_{\mathcal{O}}(\lambda_m e^{-t}) \lambda_m e^{-t}= \mathcal{L}^{-1}\left[ \lambda_m^{-n} tr (\rho_A^n \mathcal{O}) \right](t).
\eea
The one-point functions $P_{\mathcal{O}}(\lambda)=\mathcal{P}_{\mathcal{O}}(t)/\lambda|_{t\to \log(\lambda_m/\lambda)}$. To evaluate them one needs $tr(\rho_A^n \mathcal{O})$.  
In the path integral formalism  $tr(\rho_A^n)=Z_n(A)/Z_1^n$, where $Z_n(A)$ is the path integral on the n-sheeted surface $\mathcal{R}_n$. By the definition of the one-point function of $\mathcal{O}$ on the surface $\mathcal{R}_n$ we have
\bea \label{key1}
\langle \mathcal{O} \rangle_{\mathcal{R}_n}:= \frac{\int_{\mathcal{R}_n}D\phi \mathcal{O}e^{-I_E}}{\int_{\mathcal{R}_n}D\phi e^{-I_E}}=\frac{tr(\rho_A^n \mathcal{O})}{tr\rho_A^n}.
\eea
For a single interval in some special states of 2D CFTs one may calculate $\langle \mathcal{O}\rangle_{\mathcal{R}_n}$.  We will show some examples later.  In the limit $n\to \infty$ the domain contribution to $tr(\rho_A^n \mathcal{O})$ is given by 
\bea \lambda_m^n \sum_{m_i}~_A\langle \lambda_{m_i}| \mathcal{O} |\lambda_{m_i}\rangle_{A} \delta(\lambda_{m_i}-\lambda_m).
\eea
By using (\ref{key1}) we find
\bea\label{maxexpectation}
\sum_{m_i}~_A\langle \lambda_{m_i}| \mathcal{O} |\lambda_{m_i}\rangle_{A} \delta(\lambda_{m_i}-\lambda_m)=\lim_{n\to\infty} \langle \mathcal{O}\rangle_{\mathcal{R}_n}.
\eea
To calculate the one-point functions in other eigenstates one should evaluate the inverse Laplace transformation.  We  define the average one-point functions
\bea
\mathcal{\bar P}_{\mathcal{O}}(t):=\frac{\mPO(t)}{\mathcal{P}(t)}.
\eea

\subsection{Example: one interval in the vacuum state of 2D CFTs}
\subsubsection{Density of eigestates}
Once  R\'enyi entropy of all the index $n$ are known we could obtain the density of eigenstates by using (\ref{distributiont}). For 2D CFTs there are several known  examples including a single interval with length $l$ in the vacuum state  or finite temperature and half infinite line in the regularized boundary states \cite{Cardy:2016fqc}. For all these examples the modular Hamiltonian $H_A$ can be written as local integral over  energy density in the subsystem $A$,
\bea\label{modularH}
H_A=\int_A dx f(x)T_{00}(x),
\eea  
where $f(x)$ depends on the situation we are considering.  For all the examples the R\'enyi entropy is like the form
\bea\label{renyientropyoneinterval}
S^{(n)}_A=(1+\frac{1}{n})b.
\eea
 Let's consider some examples. $b=\frac{c}{6} \log \frac{l}{\epsilon}$ if the size of system is infinite, where $\epsilon$ is the UV cut-off. $b=\frac{c}{12}\log \frac{2l}{\epsilon}$ if $A$ is the interval at the end of a semi-infinite line. $b=\frac{c}{6} \log \left[\frac{\beta}{\pi \epsilon}\sinh(\frac{\pi l}{\beta})\right]$ if the system is in the  canonical ensemble thermal state with inverse temperature $\beta$. \\
By using (\ref{distributiont}) and (\ref{renyientropyoneinterval}) we have
\bea
\mathcal{P}(t)=\delta(t)+ \frac{\sqrt{b}I_1(2\sqrt{bt})}{\sqrt{t}}H(t),
\eea
where $H(t)$ is the Heaviside step function, $I_n(x)$ is the modified Bessel function of the first kind. 
One can check it satisfies the normalization condition 
\bea
tr \rho_A= \int_0^{+\infty}\mathcal{P}(t) \lambda_m e^{-t} dt =1,
\eea
and the EE
\bea
S_A= -\int_0^{+\infty}dt (\lambda_m e^{-t})\log (\lambda_m e^{-t}) \mathcal{P}(t)=2b.
\eea
In this paper we are interested in the behavior of $\mathcal{P}(t)$ in the large $c$ limit. 
By using the fact $I_n(x)\simeq \frac{e^x}{\sqrt{2\pi x}}$ in the large $x$ limit,  $\mathcal{P}(t)$ can be approximated by
\bea\label{Ptoneintervalapproximation}
\mathcal{P}(t)\simeq \delta(t)+\frac{b e^{2\sqrt{bt}}}{\sqrt{4\pi}(bt)^{3/4}}H(t).
\eea
For $t=b$ or equally $\lambda=\lambda_0:=e^{-2b}$ we find $\mathcal{P}(t)\sim e^{2b}$.  
 The entropy of the state $\rho_{A,m}$ is  $S_{A,m}\simeq 2b+o(c)$
Here we define the  microcanonical ensemble state 
\bea\label{microcanonical}
\rho_{A,m}=\frac{1}{\mathcal{P}(t)}\sum_i |\lambda_i \rangle_A ~_A\langle \lambda_i|\delta(t_i-t),
\eea
with $t_i=-\log (\lambda_i/\lambda_m)$. The above results show the EE of the microcanonical ensemble state with $t=b$ is equal to the EE of $A$ at the leading order of $c$.  This motivates us to propose the microcanonical ensemble state $\rho_{A,m}$ with $t=b$ can be an approximate state of $\rho_{A}$.

\subsubsection{One-point functions of quasi-primary operators in the  vacuum conformal family}\label{Example}
The quasi-primary operators in the vacuum conformal family include the stress energy tensor $T(w)$ with conformal dimension $h_T=2$, $\mathcal{A}:=(TT)-\frac{3}{10} \partial^2T$  with conformal dimension $h_{\mA}=4$ and so on. In this section for simplicity we only consider $T$ and $\mA$. One could refer to \cite{Chen:2013kpa}\cite{Chen:2013dxa} for more details on these operators. \\
By the conformal transformation $z=f(w)=\left(\frac{w+R}{w-R}\right)^{1/n}$ the n-sheet surface $\mathcal{R}_n$ is mapped to complex $z$-plane. One may get one-point function of $T$ and $\mA$ in $\mathcal{R}_n$ 
\bea
\langle T\rangle_{\mathcal{R}_n}=\frac{c}{6}\frac{n^2-1}{n^2}\frac{ R^2}{ \left(R^2-w^2\right)^2},
\quad \text{and}\quad 
\langle \mA\rangle_{\mathcal{R}_n}=\frac{c (5 c+22)}{180}\frac{ \left(n^2-1\right)^2}{n^4}\frac{ R^4}{ \left(R^2-w^2\right)^4}
\eea
Taking $\langle T\rangle_{\mathcal{R}_n}$ into (\ref{key0}) and (\ref{key1}) we get
\bea
\mathcal{P}_{T}(t)=\frac{cR^2}{6(R^2-w^2)^2} \delta(t)+\frac{cR^2}{6(R^2-w^2)^2} \frac{\sqrt{b}I_1(2\sqrt{bt})}{\sqrt{t}}\left(1-\frac{t}{b} \right)H(t).
\eea
The first term gives the contribution from  the maximal eigenvalue $\lambda_m$, which is consistent with (\ref{maxexpectation}).  For $t\ne 0$ or equally $\lambda\ne \lambda_m$  the average one-point functions of $T$ is
\bea\label{micT}
\mathcal{\bar P}_T(t)=\frac{cR^2}{6(R^2-w^2)^2} \left(1-\frac{t}{b} \right).
\eea 
Similarly, we can derive 
\bea
\mathcal{P}_{\mA}(t)= \frac{ c (5 c+22)R^4}{180 \left(R^2-w^2\right)^4} \delta(t) + \frac{ c (5 c+22)R^4}{180 \left(R^2-w^2\right)^4}\frac{\sqrt{b}I_1(2\sqrt{bt})}{\sqrt{t}} \left[1-2 \frac{t}{b}+\frac{t^2}{b^2}\frac{I_3(2\sqrt{bt})}{I_1(2\sqrt{bt})}\right]H(t). 
\eea
For $t\ne 0$ the average one-point function is
\bea
\mathcal{\bar P}_{\mA}(t)= \frac{ c (5 c+22)R^4}{180 \left(R^2-w^2\right)^4} \left[1-2 \frac{t}{b}+\frac{t^2}{b^2}\frac{I_3(2\sqrt{bt})}{I_1(2\sqrt{bt})}\right].
\eea
Note that  in the limit $c\to \infty$ the above expression can be approximated by
\bea\label{micA}
\mathcal{\bar P}_{\mA}(t)\simeq \frac{ c (5 c+22)R^4}{180 \left(R^2-w^2\right)^4} \left(1-\frac{t}{b}\right)^2
\eea
by using the fact $I_3(2\sqrt{bt})/I_1(2\sqrt{bt})\to 1$. \\
Both of the two examples show the average one-point function $\mathcal{\bar P}(t)$ is divergent at $R$ and $-R$, which are the boundary of the subsystem $A$.  Actually this should be a general result for any operators.   To obtain a well defined eigenstates and eigenvalues of $H_A$ we should make some a UV cut-off $\epsilon$ near the ending point of subsystem $A$\cite{Cardy:2016fqc}. The divergence of one-point functions is closely related to the regularization.  In the large $c$ limit  the one-point functions are vanishing in the eigenstates $t_0=b$ or $\lambda_0=\lambda_m^2$. This is consistent with our expectation that the microcanonical ensemble state $\rho_{A,m}$ (\ref{microcanonical}) can approximate $\rho_A$ since $tr(\rho_A T), tr(\rho_A\mathcal{A})=0$.
\subsubsection{General operators}\label{generalonepoint}
In Appendix.\ref{MoreExample} we calculate more examples of the one-point functions of quasi-parimary operator in vacuum conformal family.  All these examples show $\mathcal{P}_{\mathcal{O}}(t)$ is vanishing for $t=b$ in the large $c$ limit. In this section we would like to show the conclusion is true for any quasi-primary operators in vacuum family. \\
Under the conformal transformation $z=f(w)$ a general quasi-parimary operator $\mathcal{X}(w)$ with conformal dimension $h_{\mathcal{X}}$ would transform as the following,
\bea
\mathcal{X}(w)=f'(w)^{h_{\mathcal{X}}}\mX (z)+\sum_{\mY, p}F_{\mY,p}[f(w)]\partial^p \mY[f(w)]+S[f(w)],
\eea  
where the sum over $\mY$ includes all the quasi-primary operators with conformal dimension $h_\mY\le h_\mX$, $F_{\mY,p}[f(w)]$ and $S[f(w)]$ are functionals of $f(w)$ and its derivatives.  By the conformal map $f(w)=\left(\frac{w+R}{w-R}\right)^{1/n}$ the image of n-sheeted surface $\mathcal{R}_n$ is complex $z$-plane $\mathcal{C}$. By symmetry we have $\langle \mY(z)\rangle_{\mathcal{C}}=0$. This means $\langle \mX(w)\rangle_{\mathcal{R}_n}= S[f(w)]$.  Moreover, $S[f(w)]$ are composed by the Schwarzian derivative $s(w)$ and its derivatives, $s'(w), s''(w)$ and so on.  By the definition of $s(w)$ we have
\bea
s(w)=\frac{ \left(n^2-1\right)}{n^2}\frac{2 R^2}{\left(R^2-w^2\right)^2}.
\eea
By induction one can expand $S[f(w)]$ as 
\bea
S[f(w)]=\sum_{I=1}^{h_{\mX}/2} S_I \left(\frac{n^2-1}{n^2}\right)^{I},
\eea
where $S_I$ don't depend on $n$. The lower bound of the sum is not $0$ since we expect $S[f(w)]$ is vanishing in the limit $n\to 1$.
By using (\ref{key0}) we can derive the one-point functions of $\mX$,
\bea
\mathcal{P}_\mX(t)=\frac{1}{2\pi i}\int_{\gamma-i\infty}^{\gamma+i\infty} dn e^{nt+\frac{b}{n}}S[f(w)].
\eea
Since the inverse Laplace transformation is linear, to obtain the above result we only need to evaluate the $I$-th term 
\bea
\mathcal{P}_{\mX,I}(t):=\frac{1}{2\pi i}\int_{\gamma-i\infty}^{\gamma+i\infty} dn e^{nt+\frac{b}{n}}\left(\frac{n^2-1}{n^2}\right)^I=\frac{1}{2\pi i}\int_{\gamma-i\infty}^{\gamma+i\infty} dn e^{nt+\frac{b}{n}+I \log(\frac{n^2-1}{n^2})}.
\eea
For a given $I$ one could analytically calculate the integration by making derivatives on $\mathcal{P}(t)$  with respect to $b$. However, it is not easy to find a formula for any $I$. We assume $t\sim c$ which is very large. The above integral is dominated  by a saddle point with
\bea
n\simeq \sqrt{\frac{b}{t}}+o(c^0).
\eea
So the saddle point approximation of the integral is 
\bea\label{onepointgeneralX}
\mathcal{P}_{\mX,I}(t)\simeq \sqrt{\frac{b}{t}}e^{2\sqrt{b t}}\left(1-\frac{t}{b} \right)^I.
\eea
We get the one-point function
\bea
\mathcal{P}_\mX(t)=\sum_{I=1}^{h_\mX/2} S_I \mathcal{P}_{\mX,I}(t).
\eea
For $t=b$ we conclude that the one-point functions $\mathcal{P}_\mX(t)$ is vanishing for any quasi-primary operators in the vacuum conformal family. 

\subsubsection{Modular Hamiltonian }\label{ModularHamiltoniansection}
For the examples we consider the  modular Hamiltonian can be written as a local integral over energy density in the subsystem $A$ (\ref{modularH}).  One could get the function $f(x)$ by the method used in \cite{Cardy:2016fqc}. Let's consider $A$ to be a single interval on infinite line.  The modular Hamiltonian is given by
\bea\label{Modularone}
H_A= -\int_{-R}^{R}dw \frac{R^2-w^2}{2R}T(w) - \int_{-R}^{R}d\bar w \frac{R^2-\bar w^2}{2R} \bar T(\bar w).
\eea
We would like to calculate 
\bea
\mathcal{ P}_{H_A}(t):=\sum_i ~_A\langle \lambda_i| H_A|\lambda_i\rangle_A \delta(t_i -t) ,
\eea
by using (\ref{micT}). Taking (\ref{micT}) and  (\ref{Modularone}) into  the above equation,  we find the integral is divergent near the boundary point of $A$. Therefore, we should make some regularization. \\
To avoid sharp bipartition of Hilbert space $\mathcal{H}$ into $\mathcal{H}_A$ and $\mathcal{H}_{\bar A}$, one should put a UV cut-off around the boundary of $A$ and $\bar A$.  In the Euclidean path integral expression  this is done by introducing a slice around the boundary point of $A$ and $\bar A$.  The effect on the modular Hamiltonian is changing the integral region to be $[-R+\epsilon,R-\epsilon]$.\\
For $t\ne 0$ by using (\ref{micT}) and (\ref{Modularone})  we have
\bea\label{Hamiltonianaverage}
\mathcal{ P}_{H_A}(t)=-\int_{-R+\epsilon}^{R-\epsilon}dw  \frac{R^2-w^2}{2R}\mathcal{ P}_T(t)-\int_{-R+\epsilon}^{R-\epsilon}d\bar w  \frac{R^2-\bar w^2}{2R} \mathcal{ P}_{\bar T}(t)=(t-b)\mathcal{P}(t),
\eea
 where we have used $b=\frac{c}{6}\log\frac{2R}{\epsilon}$.  For $t=0$ by using (\ref{maxexpectation}) we have
\bea
\mathcal{ P}_{H_A}(t=0) =b.
\eea 
With this we can calculate $tr e^{-H_A}$ by
\bea
tr(e^{-H_A}) =e^{-b} +\int_0^{+\infty}dt e^{b-t}\mathcal{ P}(t)=e^{2b},
\eea
where the first term comes from the contribution of $t=0$. Therefore, the normalized reduced density matrix $\rho_A$ is
\bea
\rho_A= e^{-H_A-2b}.
\eea
Notice that the spectrum of $H_A+2b $ is $t+b$, which is positive. So the operator $H_A+2b$ is  a positive operator. \\
One may construct other operators like the modular Hamiltonian, such as
\bea
Q_\mA=\int_{-R}^{R}dw \frac{(R^2-w^2)^3}{R^3}\mA(w)+\int_{-R}^{R}d\bar w \frac{(R^2-\bar w^2)^3}{R^3}\bar{\mA}(\bar w).
\eea
The expectation value of this operator in the microcanonical ensemble state (\ref{microcanonical}) is 
\bea\label{Aaverage}
\mathcal{P}_{Q_\mA}(t)\simeq \frac{bc}{3}(1-\frac{t}{b})^2 \mathcal{P}(t),
\eea
where we have assumed the large $c$ limit and used (\ref{micA}).  It seems the operators $Q_A$ and $H_A$ don't have common eigenstates $|\lambda_i\rangle_A$. However, $Q_A$ still show very similar behavior as the modular Hamiltonian. One also could construct  other similar operators by $\mathcal{B}$, $\mD$, etc.   It is interesting to explore these kinds of operators in the future. 

\subsubsection{Two-point functions}\label{twopointsection}
 The one-point functions of primary operators  are vanishing for a single interval in infinite line because $\langle O\rangle_{\mathcal{R}_n}=0$. For the interval at the end of a semi-infinite line or in the regularized boundary state the one-point functions are non-vanishing since the one point function in these two states can be associated with two-point functions on the complex plane. In this section we would like to discuss the two-point functions as probes to detect the microcanonical state $\rho_{A,m}$.\\
The two-point functions are defined as
\bea
\mathcal{P}_{\mathcal{O}\mathcal{O}}(t):= \sum_i ~_A\langle \lambda_i | \mathcal{O}(x)\mathcal{O}(y)|\lambda_i\rangle_A\delta(t_i-t).
\eea
The average two-point function is $\mathcal{\bar P}_{\mathcal{O}\mathcal{O}}(t):= \mathcal{P}_{\mathcal{O}\mathcal{O}}(t)/\mathcal{P}(t)$.
Without loss of generality we choose $0<x<R$ and $y=-x$. To calculate the two-point functions we need
\bea
\langle \mathcal{O}(x)\mathcal{O} (-x) \rangle_{\mathcal{R}_n}=f'(x)^{h_{\mathcal{O}}} f'(-x)^{h_{\mathcal{O}}} \left( f(x)-f(-x)\right)^{-2h_{\mO}}=\left( \frac{2R}{n(R^2-x^2)}\right)^{2h_\mO} \left( \xi^{-\frac{1}{2n}}-\xi^{\frac{1}{2n}}\right)^{-2h_\mO},
\eea
where $\xi:= \left(\frac{R-x}{R+x}\right)^2$.  By using (\ref{key0}) and (\ref{key1}) and the above result we can obtain 
\bea
\mathcal{P}_{\mathcal{O}\mathcal{O}}(t)=\frac{1}{2\pi i}\int_{\gamma-i\infty}^{\gamma+i \infty}dn e^{nt +\frac{b}{n}}\langle \mathcal{O}(x)\mathcal{O} (-x) \rangle_{\mathcal{R}_n}. 
\eea
We cannot find an analytical result of the inverse Laplace transformation. If the distance between the two operators is small, i.e., $x\ll R$, we have $\xi \sim 1$ and 
\bea
\langle \mathcal{O}(x)\mathcal{O} (-x) \rangle_{\mathcal{R}_n}=(2x)^{-2h_\mO}+\frac{h_\mO \left(n^2-1\right) (2 x)^{2-2 h_\mO}}{3 n^2 R^2}+o(x^{2-2h_\mO}).
\eea
For $t\ne 0$ this gives 
\bea\label{twopointinvacuumnear}
\mathcal{P}_{\mathcal{O}\mathcal{O}}(t)=\mathcal{P}(t) (2x)^{-2h_\mO}\left(1+\frac{h_\mO  (2 x)^{2}}{3 R^2}(1-\frac{t}{b})+o(x^{2})\right).
\eea
Here we only list the results upto $O(x^2)$. In Appendix.\ref{higher} we calculate the two-point functions upto $O(x^6)$. An important feature of the results upto $O(x^6)$ is 
\bea
\mathcal{\bar P}_{\mathcal{O}\mathcal{O}}(t=b)=(2x)^{-2h_\mO}+o(x^6),
\eea
in the large $c$ limit,
which is consistent with the two point correlation function in vacuum state $\langle O(x)O(-x)\rangle=(2x)^{-2h_\mO}$ upto $O(x^6)$. This further supports our expectation that the mircocanonical state $\rho_{A,m}$ can approximate the reduced density state $\rho_A$ in the large $c$ limit.  \\
If the distance between two operators is large, i.e., $x\sim R$,  we have $\xi\sim 0$ and
\bea
\langle \mathcal{O}(x)\mathcal{O} (-x) \rangle_{\mathcal{R}_n}\simeq  \left( \frac{2R}{n(R^2-x^2)}\right)^{2h_\mO} \xi^{\frac{h_\mO}{n}}.
\eea 
For $t\ne 0$ we get
\bea
\mathcal{P}_{\mathcal{O}\mathcal{O}}(t)=\left(\frac{2R}{R^2-x^2}\right)^{2h_\mO}\left(\frac{b+h_\mO\log\frac{R-x}{R+x} }{t}\right)^{\frac{1-2h_\mO}{2}}I_{2h-1}(2\sqrt{t(b+h_\mO\log \frac{R-x}{R+x})}).
\eea
Even if taking $t=b$ we find the two-point functions are not same as $\langle \mathcal{O}(x)\mathcal{O}(-x)\rangle$. 
\subsection{Indistinguishability of $\rho_{A,m}$ from $\rho_A$}
In the previous sections we calculate the one-point functions and two-point functions in the microcanonical state $\rho_{A,m}$. If taking the large $c$ limit, we find the one-point functions in the state $\rho_{A,m}$  with $t=b$ is vanishing. The two-point correlation functions of primary operators are equal to the ones in $\rho_A$ upto $O(x^6)$, where $2x$ is the distance of the two operators. These are consistent with our expectation that $\rho_{A,m}$ with $t=b$ can be seen as an approximate state of $\rho_{A}$ in the large $c$ limit. However,   if the distance between the two operators are large $x\sim R$,  the two-point functions $\mathcal{P}_{\mO\mO}(t)$ are different from the ones in $\rho_A$. Therefore, a reasonable explanation of these results is $\rho_{A,m}$ is indistinguishable from $\rho_A$ in large $c$ limit only if the probes are located in a small region comparing with $R$, that is $T/R\ll 1$. One could define a reduced density matrix $\rho_{A_3,m}:=tr_{A_1A_2} \rho_{A,m}$. In the case  $T/R\ll 1$ one could show that the distance between $\rho_{A_3,m}$ and $\rho_{A_3}$ would approach to $0$ in the large $c$ limit , i.e.,
\bea
\lim_{c\to \infty}d(\rho_{A_3,m},\rho_{A_3}) \to 0,
\eea
where $d(\rho,\sigma)$ are the quantities that characterize the distance between two density matrices, such as trace distance, relative entropy, etc. There are some studies on relative entropy and trace distance in CFTs, see\cite{Zhang:2019wqo}-\cite{Sarosi:2016atx}. Here we will not dicuss the details of the distance, but one could calculate the trace distance and relative entropy by using our results in previous sections. \\
However, in the case $T/R \sim 1$ the two states $\rho_{A,m}$ and $\rho_{A}$ are distinguishable, for example one could use the two-point function with distance $2x$ as a probe. This phenomenon also appears in the discussion of the canonical ensemble thermal state and the microcanonical state. One could show the reduced density matrices of a small subsystem are indistinguishable for the two ensemble states ( see the supplemental material of \cite{Guo:2018djz}). However, for large subsystem there exists many probes to distinguish them. One of them is the R\'enyi entropy \cite{Dong:2018lsk}\cite{Guo:2018fye}. Here we can also use the R\'enyi entropy to distinguish $\rho_{A,m}$ from $\rho_{A}$. Specially, for the system $A$ their R\'enyi entropies are different,
\bea
S^{(n)}(\rho_{A,m})=2b,\quad S^{(n)}(\rho_{A})=(1+\frac{1}{n})b.
\eea
In a summary, $\rho_{A_3,m}$ and $\rho_{A_3}$ are indistinguishable  if the size of $A_3$ is small enough, but they are distinguishable if the size of $A_3$ is large.

\section{Entanglement spectrum of  geometric states} 
In this section we will generalize the results of the single interval example to  arbitrary geometric states.  
\subsection{Two intervals in vacuum state}\label{twointervalinvacuum}
Before we discuss the general case, let's see a non-trivial  example.  Without loss of generality we could consider the two intervals  $A_1$ and  $A_2$ as mentioned in the introduction.  Let's define the R\'enyi mutual information
\bea
I^{(n)}(\rho_{A_1A_2}):= S^{(n)}(\rho_{A_1})+S^{(n)}(\rho_{A_2})-S^{(n)}(\rho_{A_1A_2}).
\eea
Take $n\to 1$ we get the mutual information $I(\rho_{A_1A_2})$. By conformal symmetry one can show the R\'enyi mutual information $I^{(n)}(\rho_{A_1A_2})$ only depends on the cross ratio $\eta:= \left(\frac{1-T/R}{1+T/R}\right)^2$. Let's denote $ I^{(n)}(\rho_{A_1A_2})$ to be $I_n(\eta)$. 
For $T> (3-2\sqrt{2})R$ we have $I(\rho_{A_1A_2})=O(c^0)$, which means
\bea
\rho_{A_1A_2}=\rho_{A_1}\otimes \rho_{A_2}+\delta\rho,
\eea
where $\delta \rho$ is taken as a small perturbation in the large $c$ limit.  To derive the density of eigenstates of $\rho_{A_1A_2}$ one needs the R\'enyi entropy for any index $n$ as we have done for the single interval case. Unfortunately, we have no exact results of the R\'enyi entropy. But if the distance between the two interval $A_1$ and $A_2$ is large, one could perturbatively calculate the R\'enyi entropy by short interval method\cite{Headrick:2010zt}-\cite{Calabrese:2010he}. For $\eta \ll 1$  we have $S^{(n)}(\rho_{A_1A_2})=2(1+\frac{1}{n})b-I_n(\eta)$, where $b:=\frac{c}{6}\log (R-T)/\epsilon$  and
\bea\label{mutuallarge}
&&I_{n}(\eta)= \frac{c(n-1)(n+1)^2\eta^2}{144n^3}+\frac{c(n-1)(n+1)^2\eta^3}{144n^3}+\frac{c(n-1)(n+1)^2(1309n^4-2n^2-11)\eta^4}{207360n^7}\nonumber \\
&&\phantom{I_{n}(\eta)=}+o(\eta^4,c),
\eea
where we only keep the order of $O(c)$ results upto $\eta^4$.  One could refer to \cite{Chen:2013kpa} for the higher order terms. Note that the above results only contain the contributions from the quasi-primary operators in the vacuum conformal family. In the theory with holographic dual we expect these operators gives the main contributions to the R\'enyi entropy. \\
We can formally write $\rho_{A_1A_2}$ as
\bea
\rho_{A_1A_2}=\sum_{J} \sqrt{\Lambda_{J}}|J\rangle_{A_1A_2}~_{A_1A_2}\langle J|,
\eea
and define the density of eigenstates
\bea
P(\Lambda)=\sum_J \delta(\Lambda_J-\Lambda).
\eea
Using the expression (\ref{mutuallarge}) one can find the maximal eigenvlue  $\Lambda_m$ by
\bea
B=-\log \Lambda_m=\lim_{n\to \infty}S^{(n)}(\rho_{A_1A_2})=2b+\delta b+o(\eta^4,c),
\eea
where  $\delta b=-\frac{c \eta ^2 \left(1309 \eta ^2+1440 \eta +1440\right)}{207360}$.
By using (\ref{distributiont}) we obtain 
\bea
\mathcal{P}(t):=P(\Lambda_m e^{-t})\Lambda_m e^{-t}=  \frac{1}{2\pi i} \int_{\gamma-i\infty}^{\gamma+i\infty}dn e^{nt+n B +(1-n)S^{(n)}(\rho_{A_1A_2})}
\eea
With some calculations the integral becomes
\bea\label{Ptwointerval}
\mathcal{P}(t)= \frac{1}{2\pi i} \int_{\gamma-i\infty}^{\gamma+i\infty}dn e^{nt+\frac{2b}{n}+n \delta b -(1-n) I_{n}(\eta)}.
\eea
One could evaluate this integral by the saddle point approximation. The approximation is safe if we assume $t\sim c$. We need to solve the equation
\bea
t-\frac{2b}{n^2}+\delta b-\frac{\partial}{\partial n}[(1-n)I_n(\eta)]=0.
\eea
Since we take $\eta\ll 1$ the equation can be solved perturbatively. The solution is
\bea
&&n_0=\sqrt{\frac{2 b}{t}}-\frac{c \eta ^2 (4 b-3 t)}{576   \sqrt{2b^3 t}}-\frac{c \eta ^3 (4 b-3 t)}{576  \sqrt{2b^3 t}}-\frac{c\eta^4}{6635520   \sqrt{2b^7 t}}\nonumber \\
&&\phantom{n_0=}\times\left[80 b^2 (524 b+c)+t^2 (225 c-400 b)-72 t (b (434 b+5 c))+154 t^3\right]+o(\eta^4).
\eea
Taking the above equation back into (\ref{Ptwointerval}), we have
\bea
\mathcal{P}(t)\simeq n_0 e^{s(t)},
\eea
with
\bea
&&s(t)=2 \sqrt{2 b t}-\frac{c \eta ^2 (4 b-t) \sqrt{b t}}{288 \sqrt{2} b^2}-\frac{c \eta ^3 (4 b-t) \sqrt{b t}}{288 \sqrt{2} b^2}+\frac{c \eta ^4 \sqrt{t}}{3317760 \sqrt{2} b^{7/2}}\nonumber \\
&&\phantom{s(t)=}\times \left[80 b^2 (524 b+c)+t^2 (45 c-80 b)-24 t (b (434 b+5 c))+22 t^3\right]+o(\eta^4).
\eea
Motivated by the discussions in section.\ref{densitysingle} we would like to find a microcanonical ensemble states similar with  $\rho_{A,m}$(\ref{microcanonical}) with $t=t_0$  that can be taken as an approximate state of $\rho_{A_1A_3}$.  A necessary condition for this is the entropy of the microcanonical state $\log \mathcal{P}(t_0)$ is consistent with the EE of $\rho_{A_1A_3}$.  We find the non-trivial solution is 
\bea
t_0=2b-\delta b+o(\eta^4).
\eea
Thus we have $s(t_0)=4b+o(\eta^4)$ or
\bea
\mathcal{P}(t_0) \simeq n_0 e^{4b +o(\eta^4)}.
\eea
For $T<(3-2\sqrt{2})R$  we have $\eta\ge \frac{1}{2}$. Specially, for $T\ll (3-2\sqrt{2})R$  the perturbation results may broke down.  However, we can use the following results\cite{Headrick:2010zt},
\bea
I_n(\eta)=I_n(1-\eta)+\frac{(n+1)c}{6n}\log \frac{\eta}{1-\eta}.
\eea
This gives
\bea
S^{(n)}(\rho_{A_1A_2})=S^{(n)}(\rho_{A_1})+S^{(n)}(\rho_{A_2})-I_n(1-\eta)-\frac{(n+1)c}{6n}\log \frac{\eta}{1-\eta}.
\eea
Note that $S^{(n)}(\rho_{A_1})=S^{(n)}(\rho_{A_2})=\frac{c(n+1)}{6n}\log\frac{R-T}{\epsilon}$.  The R\'enyi entropy can be written as
\bea\label{twointervalsmall}
S^{(n)}(\rho_{A_1A_2})=(1+\frac{1}{n})b_1+(1+\frac{1}{n})b_2-I_n(1-\eta),
\eea
with $b_1=\frac{c}{6}\log\frac{2T}{\epsilon}$ and $b_2=\frac{c}{6}\log\frac{2R}{\epsilon}$. The calculations are similar as we have done for the case $T>(3-2\sqrt{2})R$ by replacing $2b$ with $b_1+b_2$ and $\eta$ with $1-\eta$.  Therefore,  $t_0$ is given by
\bea\label{twointervalb}
t_0=b_1+b_2-\delta b',
\eea
with $\delta b'=-\frac{c (1-\eta) ^2 \left(1309(1-\eta) ^2+1440(1-\eta) +1440\right)}{207360}$.
\subsection{Arbitrary geometric states}
Our previous discussions on entanglement spectrum on single interval or two intervals in vacuum states show interesting properties of the holographic theories. In this section we will generalize the discussions to arbitrary geometric states.
To do that  we need to know the scale behavior of R\'enyi entropy in the large $c$ limit. 
\subsubsection{The gravity dual of R\'enyi entropy}
  Let's first review the proposal of holographic R\'enyi entropy in \cite{Dong:2016fnf}. The holographic R\'enyi entropy also follow an area law like the RT formula.  For a subsystem $A$ it is given by
\bea
n^2 \partial_n\left(\frac{n-1}{n}S^{(n)}(\rho_{A})\right)=\frac{\text{Area}(\mathcal{B}_n)}{4G},
\eea
where $\mathcal{B}_n $ denotes a bulk codimension-2 cosmic brane homologous to the boundary region $A$ . The tension $T_n$ of  $\mathcal{B}_n$ is associated with the R\'enyi index $n$ by
\bea
T_n=\frac{n-1}{4nG}.
\eea
One may obtain the bulk geometry as well as the area of cosmic brane $\mathcal{B}_n$ by solving the Einstein equation with the Euclidean action $I=I_{\text{bulk}}+I_{\text{brane}}$, where $I_{\text{bulk}}$ includes the Einstein-Hilbert action and the matter field, $I_{\text{brane}}=T_n \int d^{y-1}\sqrt{g}$. In this paper we will not pursue the solutions for special cases.  What we need is  
that the metrics for any geometric states should be of order $O(G^0)$, since the action of the cosmic brane is of order $O(G^{-1})$ same as the bulk actions. Therefore, the area of $\mathcal{B}_n$ should be of order $O(G^{0})$, which means R\'enyi entropy is of $O(G^{-1})$ or $O(c)$ by $c\sim 1/G$.  The RT formula is a special case of the holographic R\'enyi entropy formula in the probe limit $n\to 1$. \\
A natural assumption is the existence of the limit $S^{\infty}:=\lim_{n\to \infty}S^{(n)}(\rho_A)$.  We further assume $S^{(n)}(\rho_A)$ can be expanded as follows,
\bea\label{ansatz}
S^{(n)}(\rho_A)=S^{\infty}+\sum_{k=1} \frac{b_k}{n^k},
\eea
where $b_k$ are parameters of $O(c)$. The above ansatz is consistent with the existence of $S^{\infty}$. One could refer to \cite{Hung:2011nu}\cite{Belin:2013dva} for more supports on the ansatz. Take the limit $n\to 1$ we get the EE
\bea
S(\rho_{A})=S^\infty+\sum_{k=1}b_k.
\eea
The  single interval and two intervals examples both satisfy the ansatz (\ref{ansatz}).  It is out of the scope of this paper to show the ansatz is true for any geomemtric states.  To approach this one needs to study more details of the solutions of  Einstein equation. 
\subsubsection{Density of eigenstates and microcanonical ensemble}
The density of eigenstates can be obtained by using (\ref{distributiont}),
\bea
\mathcal{P}(t)=\frac{1}{2\pi i}\int_{\gamma-i\infty}^{\gamma+i\infty}dn e^{s_n},
\eea
with 
\bea\label{sn}
s_n=n t+nS^{\infty} +(1-n)\left(S^{\infty}+\sum_{k=1} \frac{b_k}{n^k} \right),
\eea
where we have used the maximal eigenvalue $\lambda_m=e^{-S^\infty}$. We will take the large $c$ limit and assume $t\sim c$. Thus we could use the saddle point approximation to evaluate the inverse Lapalce transformation. We can rewrite $s_n$ as
\bea
s_n= nt +nS^\infty+(1-n)S(\rho_A)+(1-n)^2\sum_{k=1} \frac{(\sum_{k_i=0}^{k-1}n^{k_i})b_k}{n^k}.
\eea
The saddle point approximation requires the solution of the equation
\bea
\partial_n s_n =t+S^\infty-S(\rho_A)-2(1-n)\sum_{k=1} \frac{(\sum_{k_i=0}^{k-1}n^{k_i})b_k}{n^k}+(1-n)^2 \partial_n\left(\sum_{k=1} \frac{(\sum_{k_i=0}^{k-1}n^{k_i})b_k}{n^k} \right)=0.
\eea
For general $t$ it is hard to solve the above equation even if one knows the parameters $b_k$. However, a special case is at the point $t_0=S(\rho_A)-S^\infty$, for which the solution is $n=1$. Taking the solution back to $s_n$, the density of eigenstates is given by
\bea
\mathcal{P}(t_0)\simeq e^{S(\rho_{A})}.
\eea
One could construct the microcanonical ensemble states  $\rho_{A,m}$ (\ref{microcanonical}) with  $t=t_0$, the entropy of which  is equal to the EE of $A$ upto the leading order of $c$.
 This supports that the microcanonical ensemble state with can be taken as approximate state of $\rho_A$. \\
As a check of the general result, for $T<(3-2\sqrt{2})R$ in section.\ref{twointervalinvacuum} by using (\ref{twointervalsmall}) we have $S^\infty=b_1+b_2+\delta b'$ and $S(\rho_{A_1A_2})=2(b_1+b_2)$. We obtain $t_0= b_1+b_2-\delta b'$ by using the results in this section, which is consistent with (\ref{twointervalb}).

\subsubsection{Correlation functions in microcanonical ensemble state}\label{generallocaloperators}
 Generally, we will consider  $tr(\rho_{A,m}\mathcal{S})$ where $\mathcal{S}$ denotes the product of local operators, i.e.,
\bea\label{localproduct}
\mathcal{S}=\mathcal{O}(x_1)\mathcal{O}(x_2)...\mathcal{O}(x_n).
\eea
To get $tr(\rho_{A,m}\mathcal{S})$ one needs to know $\langle \mathcal{S}\rangle_{\mathcal{R}_n}$ by using (\ref{key0}) and (\ref{key1}). Let's denote the maximal distance between the boundary of $A$  and the set $\{x_1,...,x_n\}$ to be $D_m$. And denoting  $x_m$ to be the maximal distance among the set $\{x_1,...,x_n\}$. If the operators of $\mathcal{S}$ is located in a small region in $A$ and far away from the boundary, that is $x_m\ll D_m$, we expect the following expansion
\bea\label{OPES}
\langle \mathcal{S}\rangle_{R_n}=tr(\rho_A \mathcal{S})+\sum_{k=1} (1-n)^k S_k,
\eea
where $S_k$ are some functions depending on $x_1,...,x_n$. The expansion is consistent with the fact that $\lim_{n\to 1}\langle \mathcal{S}\rangle_{\mathcal{R}_n}=tr(\rho_A \mathcal{S})$.  But we don't expect this expansion is still true for $x_m\sim  D_m$. The correlation functions $\langle \mathcal{S}\rangle_{\mathcal{R}_n}$ on $\mathcal{R}_n$ is same as correlation functions with the inserting of the twist operators $\sigma_n$ which is located at the boundary of $A$. For $x_m\sim D_m$ some operators in $\mathcal{S}$ are near the boundary of $A$. Therefore, the operator product expansion (OPE) of $\mathcal{S}$ and $\sigma_n$ would give main contributions to the correlation functions. The expansion form (\ref{OPES}) may broke down. Our example in section.\ref{twopointsection} can be taken as an non-trivial support on our argument.  
Specially, the expansion (\ref{OPES}) is expected to be true for $\mathcal{S}$ being the quasi-primary operators $\mathcal{X}$ in the vacuum conformal family. \\
By using (\ref{key0}) and (\ref{key1}) we get the correlation functions in the microcanonical state 
\bea
\mathcal{P}_{\mathcal{S}}(t)=  \frac{1}{2\pi i}\int_{\gamma-i\infty}^{\gamma+i\infty}dn e^{s_n} \langle \mathcal{S} \rangle_{\mathcal{R}_n},
\eea
where $s_n$ is given by (\ref{sn}).  Since the saddle point approximation would give $n=1$ if $t=t_0=S(\rho_A)-S^\infty$, the only survive term in (\ref{OPES}) is $tr(\rho_A \mathcal{S})$. This gives
\bea
\mathcal{P}_{\mathcal{S}}(t=t_0)\simeq \mathcal{P}(t_0) tr(\rho_A \mathcal{S}),
\eea
or 
\bea\label{resultsgeneral}
\mathcal{\bar P}_{\mathcal{S}}(t=t_0):= \frac{\mathcal{P}_{\mathcal{S}}(t=t_0)}{\mathcal{P}(t_0)}\simeq  tr(\rho_A \mathcal{S}).
\eea
The physical meaning underlying the results is $\rho_{A,m}$ and $\rho_A$ are indistinguishable if the probes are located in a small region $\tilde{A}$ and far away from the boundary of $A$. \\
We may define the reduced density matrix of the small region $\tilde{A}$ as $\rho_{\tilde A,m}:=tr_{A-\tilde{A}}\rho_{A,m}$ and $\rho_{\tilde{A}}:=tr_{A-\tilde{A}}\rho_A$. Specially, for arbitrary states in 2D CFT one could show this statement by using the short interval expansion of relative entropy , which can be seen as a measure of distance between two states $\rho_{\tilde{A},m}$ and $\rho_{\tilde{A}}$. The relative entropy can be expanded as powers of the length of interval $\ell$\cite{Sarosi:2016oks}\cite{Sarosi:2016atx}. We can show any order of the expansion is associated with $tr(\rho_A\mX)-tr(\rho_{A,m}\mX)$, where $\mX$ denotes the operators contained in the theory.  Our result (\ref{resultsgeneral}) shows the relative entropy is vanishing to any order of $\ell$ in the order of $O(c)$.  By using the Pinsker's inequality one can show the trace distance between the two states are also zero .\\
Of course, one can always find the probes to distinguish the two states, for example, the R\'enyi entropies of $\rho_A$ and $\rho_{A,m}$ are different. This shows their difference will appear if the probes are not limited to $\tilde{A}$.
\section{Applications of the microcanonical ensemble states}
In previous sections we discuss the microcanonical ensemble state $\rho_{A,m}$.The number of the microstates in the ensemble is  large which is of order $O(e^{c})$.  In this section we will come back to (\ref{ArakiLiebequality}) and find its constraints on the eigenstates and eigenvalues of $\rho_A$. \\
\subsection{Holevo information}\label{HHH}
Let's consider the subsystem  $A=A_1A_2A_3$ in the vacuum state of 2D CFTs. We have the reduced density matrix $\rho_{A_1A_2}$ and $\rho_{A_3}$,
\bea\label{ensemble12and3}
\rho_{A_1A_2}=\sum_i \lambda_i \rho_{i,A_1A_2},\quad \rho_{A_3}=\sum_i \lambda_i \rho_{i,A_3},
\eea
where $\rho_{i,A_1A_2}:=tr_{A_3} |\lambda_i\rangle_A~_A\langle \lambda_i|$ and $\rho_{i,A_3}:= tr_{A_1A_2} |\lambda_i\rangle_A~_A\langle \lambda_i|$.
For an ensemble of a mixed state $\rho=\sum_i p_i \rho_i$, we may define the Holevo information, 
\bea\label{Holevodef}
\chi(\rho):= S(\rho)-\sum_i p_i S(\rho_i).
\eea
The Holevo information is  an upper bound of the information that one can gain from the ensemble. It can also be used to characterize the distinguishability between the microstates $\rho_i$\cite{Bao:2017guc}.  It is obvious that the Holevo information is non-negative, i.e., $\chi(\rho)\ge 0$. It can also be shown that $\chi(\rho)\le H(p_i)=-\sum_i p_i \log p_i$. We are interested in the condition (\ref{ArakiLiebequality}), which  gives constraints on the spectrum and eigenstate of $\rho_{A_1A_2A_3}$.   We may write (\ref{ArakiLiebequality}) as
\bea
\chi(\rho_{A_1A_2})=\chi(\rho_{A_3}) +H(\lambda_i),
\eea
where $H(\lambda_i)=-\sum_i \lambda_i \log \lambda_i$, $\rho_{A_1A_2}$ and $\rho_{A_3}$ are given by the ensembles (\ref{ensemble12and3}). In the above derivation we use the fact $S(\rho_{i,A_3})=S(\rho_{i,A_1A_2})$.  Using the bound of the Holevo information we have
\bea
\chi(\rho_{A_1A_2})=H(\lambda_i) \quad \text{and}\quad \chi(\rho_{A_3})=0.
\eea
This means the microstates $\{ \rho_{i,A_3}\}$ are indistinguishable from $\rho_{A_3}$, while $\{\rho_{i,A_1A_2}\}$ can be perfectly distinguished. \\
The RT formula (\ref{RT}) and the relation (\ref{ArakiLiebequality}) only count the leading order contribution in the limit $G\to 0$. In the CFT this corresponds to the  limit $c\to \infty$ with $c\sim 1/G$.  The EE should have quantum  corrections which is of order $O(c^0)$. Therefore, the results should include higher order corrections of large $c$.  More precisely. we  have
\bea\label{holevo}
\chi(\rho_{A_1A_2})=H(\lambda_i)+O(c^0) \quad \text{and}\quad \chi(\rho_{A_3})=O(c^0).
\eea

\subsection{Operational meaning of the Holevo information}
We should stress that the formulas of Holevo information (\ref{holevo}) is equal to (\ref{ArakiLiebequality}). Holevo information has an operational meaning that is an upper bound of the accessible information from a given ensemble. The accessible information is associated with the operations or measurements on the ensemble. Generally, the measurement can be described by the so-called Positive Operator-Valued Measure (POVM) elements. The measurement includes the positive operators 
\bea
E_k:= M^\dagger_kM_k,
\eea
which satisfy the completeness condition $\sum_k E_k =I$. For the measurement on a state $\rho$ the probability of outcome $k$ is given by $p_k=tr (\rho E_k)$.The state after measurement is 
\bea
\rho_k =\frac{M_k \rho M_k^\dagger}{tr(M_k^\dagger M_k\rho)}.
\eea
In the view of algebraic quantum field theories the POVM elements belong to the  observable algebra $\mathcal{U}(A)$ associated with a region $A$. \\
For a given ensemble $\rho=\sum_{i\in X} p_i \rho_i$ we can take the index $X=\{1,2,...\}$ as a random variable with the probability $\{p_1,p_2,...\}$. With the measurement $E_Y=\{E_1,E_2,...\}$ the outcomes $Y$ denote the other random variable. We have the condition probability $p(y|x)=tr (\rho_x E_y)$ and  the joint distribution $p_{x,y}=p_x p(y|x)$. The accessible information  is given by the mutual information $I(X,Y)=H(p_x)+H(p_y)-H(p_{x,y})$ where $p_x:=\sum_{j\in Y} p_{x,y}$ and  $p_{y}:=\sum_{i\in X} p_{x,y}$. The Holevo information is a bound of the accessible information, i.e., $I(X,Y)\le \chi(\rho)$ with $\rho=\sum_{i\in X} p_i \rho_i$. With some calculations we have
\bea
H(X,Y)=\sum_{i\in X,j\in Y}p_i tr(\rho_i E_j)\log \frac{tr(\rho_i E_j)}{tr(\rho E_j)}.
\eea
Note that $H(X,Y)=0$ if and only if  $tr(\rho_i E_j)=tr(\rho E_j)$ for any $i\in X$ and $j\in Y$ that means any measurement cannot distinguish the states $\rho_i$ ($i\in X$). \\
\subsection{Holevo information $\chi(\rho_{A_3})$}\label{Holevoinformationequalshort}
Now let's turn to the reduced density matrix of a single interval in the vacuum state $ \rho_A=\sum_i \lambda_i |i\rangle_A ~_A\langle i|$ . Support the measurements are located in a small region, say $A_3$ with $T\ll R$, which means $E_j\in \mathcal{U}(A_3)$ ($j\in Y$). The mutual information  $H(X,Y)$ in the state $\rho_{A}$ is given by
\bea\label{mutualrhoA}
H(X,Y)|_{\rho_A}= \sum_{i,j}\lambda_i \langle E_j \rangle_i \log \langle E_j \rangle_i-\sum_j \langle E_j\rangle_{\rho_A} \log \langle E_j\rangle_{\rho_A},
\eea
where $\langle E_j\rangle_i:= ~_A\langle i| E_j|i\rangle_A$ and $\langle E_j\rangle_{\rho_A}:=tr(\rho_A E_j)$.  $H(X,Y)|_{\rho_A}$ is equal to the one defined by the  state $\rho_{A_3}=\sum_i \lambda_i \rho_{i,A_3}$ since $~_A\langle i| E_j|i\rangle_A=tr (\rho_{i,A_3}E_j)$.  The POVM elements $E_j$ are bounded operators which are composed by the local operators $\mathcal{O}(x)$ with $x\in A_3$. In section.\ref{generallocaloperators} we have shown the expectation value of the products of the local operators $\mathcal{S}$ (\ref{localproduct}) in the mircocanonical states with $t=t_0$ would approach to $tr(\rho_A \mathcal{S})$ in the large $c$ limit. Therefore, if $T\ll R$ we expect $tr(\rho_{A,m}E_j)\to tr(\rho_A E_j)$ in the large $c$  limit.  The first term of (\ref{mutualrhoA}) can be written as
\bea
 \sum_{i,j}\lambda_i \langle E_j \rangle_i \log \langle E_j \rangle_i=\sum_j\int_{0}^{+\infty}dt \left(\mathcal{P}(t)\lambda_me^{-t}\right)\mathcal{\bar P}_{E_j},
\eea
where 
\bea
\mathcal{\bar P}_{E_j}:= \frac{\sum_i \langle E_j\rangle_i \log \langle E_j \rangle_i \delta(t_i-t)}{\mathcal{P}(t)}.
\eea
By using (\ref{Ptoneintervalapproximation}) for $t\ne 0$ we have 
\bea
\mathcal{P}(t)\lambda_me^{-t}=\frac{b e^{-(\sqrt{b}-\sqrt{t})^2}}{\sqrt{4\pi}(bt)^{3/4}}\to \delta(t-b)
\eea
in the limit $b\to \infty$ or $c\to \infty$. Therefore, we have
\bea
 \sum_{i,j}\lambda_i \langle E_j \rangle_i \log \langle E_j \rangle_i\to \sum_j \frac{1}{\mathcal{P}(b)}\sum_i \langle E_j\rangle_i \log \langle E_j\rangle_i \delta(t_i-b).
\eea
Combination of the above results we find
\bea
H(X,Y)|_{\rho_{A}} \to H(X,Y)|_{\rho_{A,m}},
\eea
in the large $c$ or $b$ limit. One could define the Holevo information of $\rho_{A_3,m}:= \frac{1}{\mathcal{P}(b)}\sum_k \delta(t_i -b)\rho_{i,A_3}$ as $\chi(\rho_{A_3,m})$. We conclude  if $T\ll R$
\bea
\chi(\rho_{A_3})=\chi(\rho_{A_3,m}),
\eea
in the large $c$ limit. This equality gives us a way to explain the almost vanishing Holevo information of $\chi(\rho_{A_3})$ in the case $T< (3-2\sqrt{2})R$.  At present we only consider $T\ll R$. The mutual information $H(X,Y)|_{\rho_{A_3,m}}=H(X,Y)|_{\rho_{A,m}}$ is 
\bea\label{mutualinformationmicro}
H(X,Y)|_{\rho_{A_3,m}}=\frac{1}{\mathcal{P}(b)}\sum_{i,j}\langle E_j\rangle_i \log \frac{\langle E_j\rangle_i}{\langle E_j \rangle_{\rho_{A,m}}} \delta(t_i-b).
\eea
Let's denote $e_{ij}:=\frac{\langle E_j\rangle_i}{\langle E_j \rangle_{\rho_{A,m}}}$. Since  $\frac{1}{\mathcal{P}(b)}\sum_i e_{ij} \delta(t_i -b)=1$ and $\langle E_j\rangle_i \ge 0$,  it is obvious $e_{\text{max}}=\text{max}_{i,j}e_{ij}\ge 1$ and $e_{\text{min}}=\text{min}_{i,j}e_{ij}\le  1$.  The microstates in $\rho_{A_3,m}$ are eigenstates of $H_A$ with the same eigenvalues. A natural assumption is that the measurements $E_j$ cannot distinguish the microstates $|i\rangle_A $ at the leading order of $c$. For most of tensor $e_{ij}$\footnote{Here we don't need all of the tensor $e_{ij}$ satisfy (\ref{eexpansion}). Assume there exists some states with number $N$ such that $e_{ij}\sim c$. As long as the number $N\ll \mathcal{P}(b)\sim O(e^{c})$ their contributions to (\ref{mutualinformationmicro}) are exponentially suppressed. } we expect the expansion 
\bea\label{eexpansion}
\log e_{ij}=\tilde{e}_{ij}+O(1/c),
\eea 
where $\tilde{e}_{ij}$ are of order $O(c^0)$. With the ansatz we have
\bea
H(X,Y)|_{\rho_{A_3,m}}\simeq \frac{1}{\mathcal{P}(b)}\sum_{i,j}\langle E_j\rangle_i \tilde{e}_{ij}
\eea
Support $\tilde{e}_{\text{max}}:= \text{max}_{ij}|\tilde{e}_{ij}|$. We have
\bea
H(X,Y)|_{\rho_{A_3,m}}\le \tilde{e}_{\text{max}}. 
\eea
This result is consistent with $\chi(\rho_{A_3})= O(c^0)$.  
\subsection{Constraints on one-point functions of quasi-primary operators}\label{constraintonepoint}
One could calculate the EE and Holevo information for arbitrary state $\rho=\sum_i \lambda_i \rho_i$ directly by using the short interval expansion method. In the Appendix.\ref{shortinterval} we review the short expansion method. The Holevo information $\chi(\rho_A)$ is associated with the terms of the form
\bea
\langle \mX_1\rangle_\rho...\langle \mX_m\rangle_\rho-\sum_i p_i \langle \mX_1\rangle_i \langle \mX_m\rangle_i,
\eea
where $\langle \mX\rangle_i:=tr(\rho_i \mX)$. Support the length of the interval is $\ell$.  Let's consider the microcanonical ensemble state $\rho_{A,m}$.  We have shown in section.\ref{generalonepoint} the one-point function $tr(\rho_{A,m}\mX)\to 0$ in the large $c$ limit.  Therefore,  $\chi(\rho_{A,m})$ only depends on the $\langle \mX\rangle_i := ~_A\langle \lambda_i |\mX|\lambda_i\rangle_A$ with $\mX=T,\mA$ and their derivatives upto $O(\ell^{10})$. \\
We have calculated the average one-point functions of $T$ and $\mA$ in the microcanonical ensemble state $\rho_{A,m}$ in section.\ref{Example}. Let's define the functions
\bea\label{tafunction}
&&\mathsf{t}(w;\lambda_i):= ~_A\langle \lambda_i | T(w)|\lambda_i \rangle_A -\mathcal{\bar P}_T(t_i) ,\nonumber \\
&&\mathsf{a}(w;\lambda_i):=~_A\langle \lambda_i | \mA(w)|\lambda_i \rangle_A -\mathcal{\bar P}_{\mA}(t_i).
\eea
By definition  we have
\bea
\frac{1}{\mathcal{P}(t)}\sum_i \mathsf{t}(w;\lambda_i) \delta(t_i-t)=0,\quad  \frac{1}{\mathcal{P}(t)}\sum_i \mathsf{a}(w;\lambda_i) \delta(t_i-t)=0.
\eea
By using the results (\ref{Hamiltonianaverage}) we have
\bea
\int_{-R+\epsilon}^{R-\epsilon}dw\frac{R^2-w^2}{2R} ~_A\langle\lambda_i| T(w)|\lambda_i\rangle_A=\frac{1}{2}(b-t).
\eea
This gives 
 \bea
\int_{-R+\epsilon}^{R+\epsilon}dw \frac{R^2-w^2}{2R}\mathsf{t}(w;\lambda_i)=0.
\eea
 In general, we can expand $\mathsf{t}(x;\lambda_i)$ by the Legendre Polynomials as follows\footnote{We can ignore $\epsilon$ in the integration. }
\bea
\mathsf{t}(w;\lambda_i)=\sum_{j\ge 3} c_{ij} P_j(\frac{w}{R}), 
\eea
where $c_{ij}$ are constants independent with $w$. By using the short interval expansion of Holevo information (\ref{holevoshortinterval}) and the condition $\chi(\rho_{A,m})=O(c^0)$, we find the constraints on $\mathsf{t}(w;\lambda_i)$ at order of $O(\ell^4)$,
\bea\label{constraintt}
\frac{1}{\mathcal{P}(b)}\sum_i \mathsf{t}(w;\lambda_i)^2\delta(t_i-b)\lesssim O(c),
\eea
where $\lesssim$ means the left hand side is at most of order $O(c)$. A natural assumption is that most of the functions $\mathsf{t}(w;\lambda_i)$ are at most of order $O(\sqrt{c})$. One may define the maximal value of $\mathsf{t}(w;\lambda_i)$ among these functions as $\mathsf{t}(w)_{\text{max}}:=\text{max}_{\lambda_i}|\mathsf{t}(w;\lambda_i)|$. If $\mathsf{t}(w)_{\text{max}}\lesssim O(\sqrt{c})$, we would obtain (\ref{constraintt}).\\
 Taking the large $c$ limit, the term at order of $O(\ell^8)$ becomes
\bea
\frac{1}{630c^2}\frac{1}{\mathcal{P}(b)}\sum_i\left(\mathsf{a}(w;\lambda_i)-\mathsf{t}(w;\lambda_i)^2 \right)^2\delta(t_i-b).
\eea
By the similar argument as above we conclude most of $\mathsf{a}(w;\lambda_i)|_{\lambda_i=\lambda_0}$ should be at most of order $O(c)$. One 
may define   $\mathsf{a}(w)_{\text{max}}:=\text{max}_{\lambda_i}|\mathsf{a}(w;\lambda_i)|$. The assumption $\mathsf{a}
(w)_{\text{max}}\lesssim O(c)$  ensures $\chi(\rho_{A,m})\lesssim O(c^0)$ upto order $O(\ell^{8})$. \\
One could check the term of Holevo information $\chi(\rho_{A,m})\lesssim O(c^0)$ upto $O(\ell^{10})$ if $\mathsf{t}(w)_{\text{max}}\lesssim O(\sqrt{c})$ and  $\mathsf{a}(w)_{\text{max}}\lesssim O(c)$ are satisfied.\\

\section{Conclusions and Discussions}
In the context of AdS/CFT the geometric states should be the ones that are very special. They would show some well-defined and  special  properties in the semi-classical limit $G\to 0$ or $c\to \infty$.  In this paper we focus on the entanglement spectrum of the geometric states, which contain more information of the reduced density matrix $\rho_A$ than the EE.   We can use the inverse Laplace transformation with respect to the index $n$ of R\'enyi entropy.\\
A single interval $A$ in the vacuum state of 2D CFTs is the example that one can exactly get the eigenstates and eigenvalues of $\rho_A$.  By direct calculation we show there exists a mircocanonical ensemble states $\rho_{A,m}$ with $t_0=-\log \lambda_m$ can  be taken as an approximate state of $\rho_A$ in the large $c$ limit if our probes are located in a small region of $A$ and far away from the boundary. We get the conclusion by evaluating the one-point functions of primary and quasi-primary operators and two-point functions of primary operator in the mircocanonical ensemble state.  The one-point functions are always consistent with the one in $\rho_A$, that is vanishing. The two-point functions are consistent only if the distance between two operators are small. We should stress that the results are only true in the semi-classical limit $c\to\infty$. \\
For the two intervals example we evaluate the micrcocanonical ensemble state with $t_0$ (\ref{twointervalb}) by using saddle point approximation.  The parameter $b_0$ is not only associated with the length of subsystem $A_1$ and $A_2$, but also related to the cross ratio of the two intervals at the order of $O(c)$.  \\
By using the proposal of holographic R\'enyi entropy,  we generalize the results to arbitrary geometric states.  The key point is that  the holographic R\'enyi entropy is that $S^{(n)} \sim O(c)$. This permits us to use the saddle point  approximation to find the microcanonical ensemble state $\rho_{A,m}$ with the parameter $t_0$.  The solution of the saddle point approximation gives $n=1$.  The parameter $t_0$ has a simple expression $t_0=S(\rho_A)-S^{\infty}$, where $S(\rho_A)$ is the EE and $S^{\infty}$ is  $\lim_{n\to \infty}S^{(n)}$.  The entropy of  the microcanonical ensemble state $\rho_{A,m}$ with $t_0$ is equal to the EE of $A$. However, it cannot give the same R\'enyi entropy of $A$. The reason is that the microcanonical ensemble state can only be an approximate state of $\rho_A$ only if the probes are located in a small region of $A$ and far away from the boundary of $A$. This is consistent with the example of a single interval in the vacuum state.  We check this by comparing $tr( \rho_{A,m})\mathcal{S}$ with $tr(\rho_A\mathcal{S})$, where $\mathcal{S}$ (\ref{localproduct}) is product of the local operators.  A remarkable result is that  $tr( \rho_{A,m})\mathcal{S} \to tr(\rho_A\mathcal{S})$ if $\mathcal{S}$ permits the expansion as (\ref{OPES}), which is expected to be true if the local operators are located in a small region in $A$ and far away from the boundary. \\
Finally, we discuss the equality condition of the Araki-Lieb inequality (\ref{ArakiLiebequality}). The condition can be reformed as  the Holevo information, which can be taken as  an upper bound of information that one can gain by arbitrary measurements. 
To satisfy this condition we find the constraints on the expectation values of measurements and local operators.  This constraints would help us to understand more on properties of geometric states. \\
In the following we will discuss some unsolved problems that are worth to explore in the future. 
\subsection{Transition between distinguishability and indistinguishability}
As mentioned above  we expect $\rho_{A,m}$ and $\rho_{A}$ are indistinguishable at the leading order of  $c$ only if the probes are located in a small region.  If  using the R\'enyi entropy or two-point correlation functions with large distance, one would find the difference between the two states.  However, we cannot find the critical point where the transition between distinguishability and indistinguishability happens.  Let's see the two-point functions (\ref{twopointinvacuumnear})  in the microcanonical ensemble $\rho_{A,m}$. The average two-point function 
$\mathcal{\bar P}_{\mO \mO}(t) \to tr(\rho_A \mO \mO)$ if $t=b$ upto $O(x^6)$.  But when the distance $2x$ between two operators is large enough, the perturbative expansion with respect to $x$ may broke.  The difference between $\rho_A$ and $\rho_{A,m}$ with $t=b$ will appear. \\
An interesting question is whether the transition between distinguishability and indistinguishability is associated with the critical point $T=(3-2\sqrt{2})R$. For $T< (3-2\sqrt{2})R$ we have $\chi(\rho_{A_3})=O(c^0)$.  For $T>(3-2\sqrt{2})R$ we expect $\chi(\rho_{A_3})\sim O(c)$.  In section.\ref{Holevoinformationequalshort} we show the $\chi(\rho_{A_3})\simeq \chi(\rho_{A_3,m})$ with t=b if the length of $A_3$ is small enough. $\chi(\rho_{A_3,m})=O(c^0)$ means the microstates $\rho_{A_3,i}$ of the microcanoncial ensemble states are indistinguished at the leading order of $c$ , that is the expression (\ref{eexpansion}).  In section.\ref{constraintonepoint} we directly calculate the Holevo information of $\rho_{A_3,m}$ by using  short interval expansion. The almost vanishing $\chi(\rho_{A_3,m})=O(c^0)$ gives (\ref{tafunction}) with $\mathsf{t}(w;\lambda_i)\lesssim O(\sqrt{c})$ and $\mathsf{a}(w;\lambda_i)\lesssim O(c)$.  This means the difference of the expectation values of $\mathcal{X}$ in the mircrostates $|\lambda_i\rangle_A$ of $\rho_{A,m}$ is equal to $tr(\rho_{A,m} \mathcal{X})$ in the leading order of $c$, which are natural assumptions.  Our opinion is that $\chi(\rho_{A_3,m})=O(c^0)$ is a consequence of the indistinguishability of $\rho_{A_3}$  from $\rho_{A_3,m}$.  However, we cannot prove this or disprove this point at present. We will leave this to future works.

\subsection{A possible geometric explanation of Holevo information }
If a quantity can be associated with a geometric object in the bulk, we call it a geometric probe, for example the EE or R\'enyi entropy. To find these geometric probes are important since they would help us to understand more on the properties of geometric states. A natural requirement of the geometric quantity is that they should be order of $O(c)$.  In section.\ref{HHH} we have shown the relation of holographic EE (\ref{ArakiLiebequality}) in the case $T\le (3-2\sqrt{2})R$  is equal to the  conditions of  Holevo information (\ref{holevo}).  $\chi(\rho_{A_3})$ is vanishing at the order of $O(c)$. However, if $T> (3-2\sqrt{2})R$, we expect $\chi(\rho_{A_3})$ should be the order of $O(c)$. For $T\sim R$ we will have $\chi(\rho_{A_3})\simeq S(\rho_A)\sim O(c)$.  As shown in \cite{Schumacher} the Holevo information is monotonically increasing, that is if $\rho_{X}=tr_Y \rho_{XY}$ we have $\chi(\rho_X)\le \chi(\rho_{XY})$. This leads to 
\bea
\partial_T \chi(\rho_{A_3}) \ge 0.
\eea 
Therefore, we expect $\chi(\rho_{A_3})$ should be a quantity that is  order of $O(c)$ and monotonically increasing with $T$ in the case $T>(3-2\sqrt{2})R$.  We should note that $\chi(\rho_{A_3})$ is not only dependent with the reduced density state $\rho_{A_3}$, but also on the spectrum decomposition of $\rho_{A}$. This is because the Holevo information depends on the ensemble. Generally, a mixed state $\rho$ can be written as different ensembles, say $\rho=\sum_i p_i \rho_i$ and $\rho=\sum_j q_j \rho'_j$. The corresponding Holevo information $\chi(\sum_i p_i\rho_i)$ is different from $\chi(\sum_j q_j \rho'_j)$ in general. Our discussion is based on the fixed  ensemble  $\rho_A=\sum_i \lambda_i |\lambda_i\rangle_A ~_A\langle \lambda_i |$.  Therefore, the Holevo information $\chi(\rho_{A_3})$ should depend on the eigenvalues and eigenstates of $\rho_A$. In some sense $\chi(\rho_{A_3})$ contain more information of the state $\rho_A$ than the EE, which is only the trace of the eigenvalues.
According to the so-called subregion/subregion duality in AdS/CFT\cite{Czech:2012bh}-\cite{Dong:2016eik}, the bulk region surround by the RT surface $\gamma_A$ and the boundary region $A$ , named entanglement wedge, is expected to be dual to the reduced density matrix $\rho_A$. $\chi(\rho_{A_3})$ and $\chi(\rho_{A_1A_2})$ do extract some information of $\rho_A$. If $\chi(\rho_{A_3})$ has the geometric dual, then the geoemtric object should be in the region inside the entanglement wedge.\\
However, to get the exact dual of one quantity one should be able to evaluate the quantity both in bulk and the boundary CFTs. One could also make a guess by comparing the properties on both sides like the one that is done in the proposal of holographic entanglement of purification\cite{Takayanagi:2017knl}\cite{Nguyen:2017yqw}. In the paper we are not trying to conjecture the duality of $\chi(\rho_{A_3})$.  \\
Let's comment more on the ensemble dependence of $\chi(\rho_A)$. In \cite{Guo:2020rwj} we show the $\chi(\rho_{A_3})=O(c^0)$ (by the notation of the present paper) for the ensemble of $\rho_{A_3}$ that seems not related to the one we used here.  In that paper we consider the ensemble $\rho_{A}=\sum_k p_k |\psi\rangle_k ~_k\langle \psi| $ that would make the entropy $\sum_k p_kS(\tilde{\rho}_{k,A_3})$ minimal, where $\tilde{\rho}_{k,A_3}:=tr_{A_1A_2}|\psi\rangle_k ~_k\langle \psi| $. The entropy actually is defined as the entanglement of formation (EoF), which characterizes the correlation between $A_1$ and $A_2$. By using the Koashi-Winter relation we find that $\chi(\sum_k p_k \tilde{\rho}_{k,A_3})$ is also the order of $O(c^0)$. It is interesting to study  the relation between the ensemble $\rho_A=\sum_k p_k |\psi\rangle_k ~_k\langle \psi|$ and the spectrum decomposition (\ref{spectrumdecompositionintroduce}) in the near future.
\subsection{On construction of new geometric states}
We only focus on the microcanonical ensemble state $\rho_{A,m}$ with $t=t_0=S(\rho_A)-S^\infty$. We could construct a pure state  
\bea
|\psi\rangle_0= \frac{1}{\sqrt{\mathcal{P}(t_0)}} \sum_i|\lambda_i \rangle_A\otimes |\bar \lambda_i\rangle_{\bar A}\delta(t_i-t_0).
\eea
Of courese, the state $|\psi\rangle_0$ is  different from the vacuum state. In \cite{Bao:2019fpq}\cite{Bao:2018pvs} the authors constructed an approximate tensor networks for geometric states in AdS/CFT,  which is generally given by
\bea
|\Psi\rangle=\sum_{I=0}^{e^{O(\sqrt{S})}} \sum_{i=0}^{e^{S-\sqrt{S}}}\sqrt{\lambda_i}|I,i\rangle_A\otimes |I,i\rangle_{\bar A},
\eea 
where $S:=S(\rho_A)$ is the EE of $A$. It seems our state $|\psi\rangle_0$ is only part of the state $|\Psi\rangle$. So $|\psi\rangle_0$ and $\rho_{A,m}$ with $t=t_0$ only catch part information of the $|\Psi\rangle$ and $\rho_A$. This may be the reason why $\rho_{A,m}$ can be taken as the an approximate state $\rho_A$ only if the probes are located in  a small region of $A$  and far away from the boundary. At present we still don't know how to find the corrections of $|\psi\rangle_0$ and $\rho_{A,m}$ to make them to be better approximate states of $|\Psi\rangle$ and $\rho_A$.\\
At last, let's see the microcanonical ensemble state $\rho_{A,m}$ with $t\ne t_0$ in the single interval example.  By using (\ref{onepointgeneralX}) we see that the one-point functions of quasi-primary operators $\mX$ is non-vanishing for $t\ne b$. For example, the one-point function of $T$ and $\mA$ is given by (\ref{micT}) and (\ref{micA}).  In \cite{Guo:2018fnv} we find a series of  conditions associated with the one-point functions of quasi-primary operators for geometric states $\rho$. The first condtion is 
\bea
\langle \mA\rangle_\rho-\langle T\rangle_\rho^2\sim O(c),
\eea
if $\langle T\rangle_\rho \sim O(c)$.  One could check the state $\rho_{A,m}$ satisfies the above condition. By using the results in Appendix.\ref{MoreExample} one can further show $\rho_{A,m}$ satisfies  the conditions associated with higher order quasi-primary operators $(\mX=\mB,\mD)$. Therefore, this suggests the microcanonical ensemble state $\rho_{A,m}$ for any $t$ may be taken as the reduced density matrix of $A$ of some pure geometric  state.\\
In the references \cite{Dong:2018seb}\cite{ Akers:2018fow}(see also \cite{Dong:2019piw}) the authors discuss the so-called fixed-area states in the bulk side of AdS/CFT by using the quantum-error correcting code.  A noticable fact is the R\'enyi entropy in the fixed-area state is independent with $n$  at the leading order of $G^{-1}$ or $O(c)$ . This property is similar with our microcanonical ensemble state $\rho_{A,m}$ constructed by field theory side. It is interesting to check whether $\rho_{A,m}$ has some relations to the fixed-area state.

\section*{Acknowledgement}
I would like to thank Jiang Long for useful discussions. I am supported by the Fundamental Research Funds for the Central Universities under Grants NO.2020kfyXJJS041.

\appendix
\section{More examples of one-point functions of quasi-primary operators}\label{MoreExample}
In section.\ref{Example} we only calculate the one-point functions of $T$ and $\mA$. Here we will study more quasi-primary operators in the vacuum conformal family. There are two quasi-primary operators with conformal dimension $6$, that is
\bea
&& \mB=(\partial T\partial T)-\frac{4}{5}(\partial^2TT)-\frac{1}{42}\partial^4T, \nonumber\\
&& \mD= (T(TT))-\frac{9}{10}(\partial^2TT)-\frac{1}{28}\partial^4 T + \frac{93}{70c+29} \mB.
\eea
By using the same conformal map and the transformation law of $\mB$ and $\mD$ we have
\bea
\langle \mB\rangle_{\mathcal{R}_n}=\frac{ 2 c \left(n^2-1\right)^2 \left(2 (35 c+61) n^2-93\right)}{1575 n^6 }\frac{R^6}{ \left(R^2-w^2\right)^6},
\eea
\bea
\langle \mD\rangle_{\mathcal{R}_n}=\frac{ c (2 c-1) (5 c+22) (7 c+68) \left(n^2-1\right)^3}{216 (70 c+29) n^6}\frac{R^6}{ \left(R^2-w^2\right)^6}.
\eea
With the above results we can calculate the one-point functions 
\bea
&&\mathcal{P}_{\mB}(t)=\frac{2 c (70 c+122)}{1575}\frac{ R^6}{ \left(R^2-w^2\right)^6}\delta(t)+ \frac{2 c  }{1575 b^3 }\frac{R^6}{(R^2 - w^2)^6}\frac{\sqrt{b}}{\sqrt{t}I_1(2\sqrt{bt})}\Big[2 b^3 (35 c+61) \nonumber \\
&&\phantom{\mathcal{P}_{\mB}(t)=}-b^2 (140 c+337)t+14  b (5 c+22) t^2\frac{I_3(2\sqrt{bt})}{I_1(2\sqrt{bt})}-93  t^3\frac{I_5(2\sqrt{bt})}{I_1(2\sqrt{bt})} \Big],
\eea
and
\bea
&&\mathcal{P}_{\mD}(t)=\frac{c (2 c-1) (5 c+22) (7 c+68)}{216 (70 c+29)}\frac{R^6}{\left(R^2-w^2\right)^6} \delta(t)+\frac{c (2 c-1) (5 c+22) (7 c+68)}{216 (70 c+29)}\frac{R^6}{\left(R^2-w^2\right)^6} \nonumber \\
&&\phantom{\mathcal{P}_{\mD}(t)=}\times\frac{\sqrt{b} I_1\left(2 \sqrt{b t}\right)}{\sqrt{t}} \left[\left(1-\frac{3 t}{b}\right)+\frac{3t I_3\left(2 \sqrt{b t}\right)}{bI_1\left(2 \sqrt{b t}\right)}-\frac{t^2  I_5\left(2 \sqrt{b t}\right)}{b^2 I_1\left(2 \sqrt{b t}\right)}\right]\nonumber
\eea
In the large $c$ or $b$ limit one could find  an approximate expression for $\mathcal{P}_{\mD}(t)$ and $\mathcal{P}_{\mD}(t)$,
\bea
&&\mathcal{P}_{\mB}(t)\simeq \frac{2 c (70 c+122)}{1575}\frac{ R^6}{ \left(R^2-w^2\right)^6}\delta(t)+ \frac{2 c  }{1575  }\frac{R^6}{(R^2 - w^2)^6}\frac{\sqrt{b}I_1(2\sqrt{bt})}{\sqrt{t}} [2  (35 c+61)-93 \frac{t}{b}] (1-\frac{t}{b})^2,\nonumber \\
&&\mathcal{P}_\mD(t)=\frac{c (2 c-1) (5 c+22) (7 c+68)}{216 (70 c+29)}\frac{R^6}{\left(R^2-w^2\right)^6} \delta(t)+\frac{c (2 c-1) (5 c+22) (7 c+68)}{216 (70 c+29)}\frac{R^6}{\left(R^2-w^2\right)^6}\nonumber \\
&&\phantom{\mathcal{P}_\mD(t)=}\times\frac{\sqrt{b} I_1\left(2 \sqrt{b t}\right)}{\sqrt{t}} \left(1-\frac{t}{b}\right)^3.
\eea
$\mathcal{P}_\mB(t)$ and $\mathcal{P}_\mD(t)$ show similar properties as $\mathcal{P}_T(t)$ and $\mathcal{P}_{\mA}(t)$. Specially, $\mathcal{P}_\mB(t),\mathcal{P}_\mD(t) \to 0$ when $t=b$ in the large $c$ limit.  
\section{Higher order corrections of two-point functions}\label{higher}
In section.\ref{twopointsection} we calculate two-point functions $\mathcal{P}_{OO}(t)$ upto order $O(x^{2-2h_\mO})$.  Here we list more higher order terms, We have
\bea
&&\langle \mathcal{O}(x)\mathcal{O} (-x) \rangle_{\mathcal{R}_n}=(2x)^{-2h_\mO}\Big(1+\frac{h_\mO \left(n^2-1\right) (2 x)^{2}}{3 n^2 R^2}+\frac{h_\mO \left(n^2-1\right) \left(5 h_\mO n^2+4 n^2-5 h_\mO-1\right) (2 x)^4}{90 n^4 R^4}\nonumber \\
&& +\frac{h_\mO \left(n^2-1\right) \left(280 h_\mO^2 n^4-560 h_\mO^2 n^2+280 h_\mO^2+672 h_\mO n^4-840 h_\mO n^2+168 h_\mO+347 n^4-136 n^2+32\right) (2 x)^6}{45360 n^6 R^6}\nonumber \\
&&+o(x^6)\Big).\nonumber
\eea
With the above results we can derive 
\bea
&&\mathcal{P}_{\mathcal{O}\mathcal{O}}(t)= \mathcal{P}(t) (2x)^{-2h_\mO}\Big{\{}1+\frac{h_\mO  (2 x)^{2}}{3 R^2}(1-\frac{t}{b})+\Big[5 h_\mO+4-5 (2 h_\mO+1) \frac{t}{b}+   (5 h_\mO+1) \frac{I_3\left(2 \sqrt{bt}\right)}{ I_1\left(2 \sqrt{bt}\right)}\frac{ t^2}{b^2}\Big]\nonumber \\
&&\times \frac{h_\mO (2 x)^4}{90 R^4}+\Big[280 h_\mO^2+672 h_\mO+347-21 \left(40 h_\mO^2+72 h_\mO+23\right)\frac{ t}{b}+168 \left(5 h_\mO^2+6 h_\mO+1\right) \nonumber \\
&&\times \frac{I_3\left(2 \sqrt{bt}\right)}{I_1\left(2 \sqrt{bt}\right)} \left(\frac{t}{b}\right)^2-8 \left(35 h_\mO^2+21 h_\mO+4\right)\frac{I_5\left(2 \sqrt{bt}\right)}{I_1\left(2 \sqrt{bt}\right)} \left(\frac{t}{b}\right)^3\Big]\frac{h_\mO (2 x)^6}{45360 R^6}+o(x^6)\Big{\}}.
\eea
If we take the large $c$ limit of the above result, one would find
\bea
&&\mathcal{P}_{\mathcal{O}\mathcal{O}}(t)= \mathcal{P}(t) (2x)^{-2h_\mO}+\mathcal{P}(t) (2x)^{-2h_\mO}(1-\frac{t}{b})\Big[\frac{h_\mO  (2 x)^{2}}{3 R^2}+\big((5 h_\mO+4)-\frac{(5 h_\mO+1) t}{b}\big)\frac{h_\mO (2 x)^4}{90 R^4}\nonumber \\
&&\big( 347 + 672 h_\mO + 280 h_\mO^2-8  (17 + 105 h_\mO + 70 h_\mO^2)\frac{t}{b}+8 (4 + 21 h_\mO + 35 h_\mO^2)(\frac{t}{b})^2\big)\frac{h_\mO (2 x)^6}{45360 R^6)}+o(x^6)\Big].\nonumber 
\eea
The two-point functions are very simple upto $O(x^6)$ if $t=b$, that is
\bea
\mathcal{P}_{\mathcal{O}\mathcal{O}}(t=b)=\mathcal{P} (b) (2x)^{-2h_\mO}+o(x^6),
\eea 
or
\bea
\mathcal{\bar P}_{\mathcal{O}\mathcal{O}}(t=b)= (2x)^{-2h_\mO}+o(x^6).
\eea
\subsection{Review of short interval expansion}\label{shortinterval}
In this section we briefly review the short interval expansion of the EE and Holevo information. One could refer to \cite{Chen:2013dxa}\cite{Guo:2018fnv},\cite{Guo:2018djz} for more details. If we only consider the contributions from the vacuum conformal family, the EE of $A$ with length $\ell$ is given by
\bea
&&S_A=\frac{c}{6}\log \frac{\ell}{\epsilon}+\ell^2 a_T \langle T\rangle_\rho
                                  + \ell^4 a_{TT} \langle T\rangle_\rho^2
                                  + \ell^6 a_{TTT} \langle T\rangle_\rho^3 \nonumber \\
&& \phantom{S_A =}
           + \ell^8 \big( a_{\mA\mA} \langle A\rangle_\rho^2
                        + a_{TT\mA} \langle T\rangle_\rho^2 \langle \mA\rangle_\rho
                        + a_{TTTT} \langle T\rangle_\rho^4 \big)\nonumber \\
&&\phantom{S_A=}+\ell^{10} \big( a_{T\mA\mA}\langle T\rangle_\rho\langle \mA\rangle^2_\rho+a_{TTT\mA} \langle T\rangle_\rho^3 \langle \mA\rangle_\rho+a_{TTTTT} \langle T\rangle_\rho^5\big)+F(\partial)+O(\ell^{11}),
\eea
where the coefficients $a_{\mathcal{X}_1...\mathcal{X}_m}$ are constants. We list them as follows
\bea
&&a_{T}=-\frac{1}{6},\quad a_{TT}=-\frac{1}{30c},\quad a_{TTT}=-\frac{4}{315c^2},\quad a_{TT\mA}=\frac{1}{315c^2},\nonumber \\
&&a_{\mA\mA}=-\frac{1}{126c(5c+22)},\quad a_{TTTT}=-\frac{c+8}{630c^3},\quad a_{T\mA\mA}=-\frac{16}{693c^2(5c+22)},\nonumber \\
&& a_{TTT\mA}=\frac{32}{3465c^3},\quad a_{TTTTT}=-\frac{16(c+5)}{3465c^4}.
\eea
$F(\partial)$ denotes the terms with derivatives $\partial^m \langle \mathcal{X}\rangle_\rho$. For example, at order of $O(\ell^3)$ the term is $\frac{a_T}{2}\partial \langle T\rangle_\rho$. These terms are vanishing for translation invariant states. Though in our discussion $F(\partial)$ is non-vanishing, they are not important for our results in section.\ref{constraintonepoint}. \\
By the definition of Holevo information (\ref{Holevodef})  one could calculate the Holevo information of $\rho=\sum_i p_i \rho_{i}$ by short interval method. The results are given by
\bea\label{holevoshortinterval}
&&\chi(\rho_A)=\ell^4 a_{TT}\big(\langle T\rangle_\rho^2-\sum_i p_i\langle T\rangle_i^2\big) + \ell^6 a_{TTT} \big(\langle T\rangle_\rho^3-\sum_i p_i \langle T\rangle_i^3\big)\nonumber \\
&& \phantom{\chi(\rho_A)=}+\ell^8 \big[ a_{\mA\mA} \big(\langle A\rangle_\rho^2-\sum_ip_i\langle A\rangle_i^2\big) 
                        + a_{TT\mA} \big(\langle T\rangle_\rho^2 \langle \mA\rangle_\rho-\sum_i p_i\langle T\rangle_i^2 \langle \mA\rangle_i\big)\nonumber \\
&&\phantom{\chi(\rho_A)=}+ a_{TTTT} \big(\langle T\rangle_\rho^4-\sum_i \langle T\rangle_i^4\big)  \big]+\ell^{10}\big[  a_{T\mA\mA}\big(\langle T\rangle_\rho\langle \mA\rangle^2_\rho-\sum_i p_i \langle T\rangle_i\langle \mA\rangle^2_i \big)\nonumber \\
&&\phantom{\chi(\rho_A)=}+a_{TTT\mA} \big(\langle T\rangle_\rho^3 \langle \mA\rangle_\rho-\sum_ip_i \langle T\rangle_i^3 \langle \mA\rangle_i\big)+a_{TTTTT} \big(\langle T\rangle_\rho^5-\sum_ip_i\langle T\rangle_i^5\big) \big] +G(\partial)+O(\ell^{11}).\nonumber \\
~
\eea
where $G(\partial)$ denotes the terms associated with $\partial^m \langle \mX\rangle_i$ and  $\langle \mX\rangle_i:= tr( \mX \rho_{i})$ with $\mX=T,\mA$.

\end{document}